\documentclass[12pt]{article}
\def\gmmu{\gamma _{\mu}}
\def\gmnu{\gamma_{\nu}}

\def\gmf{\gamma _{5}}

\def\ll{\langle }
\def\rl{ \rangle }

\def\cmb{f_{\pi} \frac{G_{F}}{\sqrt{2}} V_{cb}V_{ud}^{*}}
\newcommand{\beq}{\begin{equation}}
\newcommand{\eeq}{\end{equation}}
\newcommand{\bea}{\begin{eqnarray}}
\newcommand{\eea}{\end{eqnarray}}

\begin{document}
\renewcommand{\thefootnote}{\fnsymbol{footnote}}
                                        \begin{titlepage}
\begin{flushright}
TECHNION-PHYS-96-14 \\
hep-ph/9605441
\end{flushright}
\vskip1.8cm
\begin{center}
{\LARGE
Regge asymptotics and color 
suppressed heavy meson decays
            \\ }
\vskip1.5cm
 {\Large Boris~Blok} 
and 
{\Large Igor~Halperin} $\footnote { Address after 
October 1, 
1996 : Physics Department, University of British Columbia, 6224 
Agriculture Road, Vancouver, BC V6T 1Z1, Canada} $ \\
\vskip0.2cm
       Technion - Israel Institute of Technology   \\
       Department of Physics  \\
       Haifa, 32000,  Israel \\
{\small e-mail addresses: 
phr34bb@vmsa.technion.ac.il \\
higor@techunix.technion.ac.il}\\

\vskip1.5cm
{\Large Abstract:\\}
\parbox[t]{\textwidth}{
We discuss a possible generation of color suppressed B-decays amplitudes 
through a soft final state interaction. 
As a typical example, we consider in detail the decay $ 
\bar{B}^{0} \rightarrow D^{0} \pi^{0} $ (and also $ \bar{B}^{0} 
\rightarrow 2 \pi^{0} $). We show that in 
the approximation of the two particle unitarity and at zero order in 
$ \alpha_{s} $ this process can be related to  
the weak decay $ \bar{B}^{0} \rightarrow D^{+} \pi^{-} $ followed by the 
strong charge exchange scattering in the Regge kinematics. We estimate the
amplitude of this process using the light cone QCD sum rule technique and 
find that it is supppressed as a power of $ 1/m_{B} $ 
in comparison to the amplitude generated by the effective non-leptonic
Hamiltonian, but remains important for the physical value of $m_{B} $ }

\vspace{1.0cm}
{\em submitted to Physics Letters B }
\end{center}
                                                \end{titlepage}

\section{Introduction}

    Understanding final state interactions (FSI) effects in weak 
decays is a challenging task 
of acute interest, both theoretical and phenomenological.
Among plenty of problems, strongly influenced by possible FSI, one can 
mention a determination of the Kabayashi-Maskawa matrix elements, weak CP
violating phases, etc. It is tremendously difficult to address the issue 
of FSI on a quantitative level. Still,
one may believe that with going to heavier (c,b) quarks  
possible FSI 
would become less and less important. In most intuitively appealing form
this expectation has been expressed by Bjorken \cite{BJ}, who viewed this 
behavior as a typical manifestation of the color transparency phenomenon
\cite{FS}.
This scenario can be explained as follows. When a virtual W-boson decays 
into a $ u \bar{d} $ pair with a low invariant mass, this pair is 
initiallly only weakly interacting with the bulk of the system as its 
color dipole moment is small. On the other hand, the pair escapes far 
away from the rest of decay products before starts to interact strongly,
as a result of the relativistic time dilation. It follows on these grounds 
that in decays like $ B \rightarrow D \pi $ FSI die off asymptotically 
as $ m_{b} \rightarrow \infty $.  In effect, weak non-leptonic decay 
amplitudes factorize into a product of hadron matrix elements; hard 
gluons, which otherwise would violate this factorization, do not manage 
to appear on time.  

     It is worth noticing that this scenario, while is very 
attractive in 
view of its simplicity and appeals to the one's intuition, is only 
qualitative and does not predict the rate of vanishing FSI corrections 
when $ m_{b} \rightarrow \infty $. Moreover, by its very appearence, this 
argumentation is likely to be applicable only to FSI proceeding via a hard 
gluons exchange. In fact, the factorization property has been proved to all 
orders of the perturbation theory in the heavy quark limit \cite{DG}
$ m_{c}, m_{b} \rightarrow \infty $ , $ m_{c}/m_{b} $ fixed. 
On the other hand, in the case of non-perturbative soft FSI the time 
dilation is not so important, and one may expect that those effects could be 
sizeable in spite of the large b-quark mass. Moreover, if initial wave 
packages hadronize without gluon FSI when their separation is not yet 
large ( $ \sim 0.3 $ fm ), then there usual hadron-hadron scattering
would be a dominant mechanism of FSI. If this kind of FSI is strong 
enough, it presumably could screen the effect of absence of perturbative 
FSI, which is due to the color transparency phenomenon.
The present paper is just an 
attempt to provide a semi-quantitative estimate of a possible role of 
hadronic FSI
corrections in non-leptonic B-decays of the type $ B \rightarrow D \pi $.
Our approach is essentially based on the existence of large parameter in 
the problem 
( $ m_{B} $ ) , and differs from more traditional approaches 
to FSI in weak decays. 

      We would like to start our presentation by explaining why FSI 
corrections are so  elusive and hard to treat in most 
non-perturbative 
methods  which have been applied to non-leptonic B-decays. As has been 
first explained in Ref.\cite{BS87}, the problem lies in the fact that 
physical amplitudes are initially extrapolated into the 
Euclidean region where a calculation is feasible, and then are analytically
continued back into the Minkowsky space-time (this is the typical
strategy e.g. in the QCD sum rule method \cite{SVZ} or lattice 
calculations). Atter the 
Wick rotation fastly oscillating pole terms (corresponding to particles 
interacting in the final state) become falling ones and die off
in the deep Euclidean region. 
On the other hand, when the amplitude is restored from a dispersion 
relation, one integrates actually over some finite region of the 
invariant mass where these oscillating terms are averaged to some 
smooth function. In effect, amplitudes are always real within these 
methods; scattering phases cannot be calculated. An analogous statement
in the lattice approach has later become known as the Maiani-Testa no-go 
theorem \cite{MT, CFMS}.  

      In this paper we try to estimate semi-quantitatively FSI 
corrections to the B-decays of the form $ B \rightarrow D \pi $.
We consider the limit of the heavy b-quark. As for the c-quark, we 
distinguish
between two different limits: (i) c-quark is heavy with $ m_{b} \gg m_{c} $
and (ii) c-quark is light, the corresponding process is $ B \rightarrow 
\pi \pi $.
To simplify the problem and avoid complications due to presence of many 
channels,
we have chosen to look for FSI effects on color suppressed B-decays
(more specifically, we consider a particular case of the $ \bar{B}^{0} 
\rightarrow D^{0} \pi^{0} $ decay) and work at zero order in $ \alpha_{s} 
$. Our approach is strongly correlated with the fact that the energy 
excess in B-decays of this type is large which, as we show, imposes a 
relevance of the Regge approach. Surprisingly, we find that though FSI 
effects
are suppressed as a power of $ 1/m_{B} $ in comparison with direct 
amplitudes, they 
remain to be of the same order of magnitude at the physical value $ m_{B} = 
5.27 \; GeV $. 
Here we have to mention 
a recent paper \cite{Don} where similar to ours ideas have been applied
to another class of B-decays. We still believe that we manage to reach
more concrete motivations and results for our particular decay mode.
We will compare our conclusions to those of Ref. \cite{Don} in the last 
section of this paper. In Sect.2 we show that in the approximation of 
two particle unitarity the decay of interest can be related to a process 
of strong charge exchange scattering in the Regge kinematics
and calculate
a discontinuity of a corresponding Feynman diagram. In Sect. 3 and 4 we 
apply the light cone QCD sum rule technique \cite{BF} (see e.g. 
\cite{BBKR} for a mini-review) to estimate parameters 
of interest. As a by-product of our study, we derive  new light cone 
sum rules for the $ B \rightarrow D $ transition form factor (Sect.3) and
$ \rho D $ coupling constant $ g_{\rho DD} $ (Sect.4) .
In Sect.5  we compare the calculated reggeon-induced amplitude to that 
proceeding directly via corresponding terms in the effective 
non-leptonic Hamiltonian. Our result is consistent with the color 
trasparency picture as long as the former is suppressed approximately
as  $ 1/m_{B} $ in respect to the latter. However, we find 
that FSI corrections are pretty large for the physical value $ m_{B} 
\simeq 5.27 \; GeV $ and constitute 40 - 100  \% of the amplitude 
of direct process. Sect.6 contains a discussion of our results.

\section{Discontinuity of the $ \bar{B} \rightarrow D^{0} \pi^{0} $ amplitude}

We consider the following simplified model for the process of interest.
Since at the zero order in $ \alpha_{s} $ the direct decay $  \bar{B} \rightarrow 
D^{0} \pi^{0} $ does not occur, we seek for a hadronic intermediate state 
which would proceed non-perturbatively into 
the given final state. The most obvious candidate is the $  D^{+} \pi^{-} $
intermediate state which is related to $  D^{0} \pi^{0} $ by the quark 
rearrangement. In what follows we restrict ourselves by this particular 
contribution only, i.e. work in the approximation of two
particle unitarity relation. Corrections due to  
multiparticle intermediate states of the form $ D^{+} + n \pi $  are 
presumably small, though accurate estimates are troublesome. Still, 
calculations within the quark gluon string model \cite{DKTZ} indicate that
the averaged pion multiplicity is low (2-3), and thus we hope that the two
particle approximation is satisfactory with an accuracy not worse than the 
factor two.   

    The discontinuity of the 
amplitude is given by the expression
 
\beq
Disc \; M(k^{2}) = \int \frac{d^{4}p}{(2 \pi)^{2}} \delta((p-k)^{2})
\delta(p^{2} - m_{D}^{2}) T(k^{2}, (k-l-p)^{2}) A_{W}(B \rightarrow D^{+} 
\pi^{-} )\; , 
\eeq
where $ T(k^{2},(k-l-p)^{2})$ is the strong charge exchange amplitude 
for the process $ D^{+}(p) + \pi^{-}(k-p) \rightarrow D^{0}(k-l) + \pi^{0}(l)
$ and $ A_{W}( B \rightarrow D \pi) $ is the weak amplitude for the 
transition $ \bar{B}^{0} \rightarrow D^{+}(p) + \pi^{-}(k-p) $. Using the 
bare weak Hamiltonian 
\beq
H_{W} = \frac{G_{F}}{\sqrt{2}} V_{cb}V^{*}_{ud} 
(\bar{c}\Gamma_{\mu}b)(\bar{d} \Gamma_{\mu}u)
\eeq
( $ \Gamma_{\mu} = \gamma_{\mu} (1- \gmf) $ ) and taking into account delta
functions in the above formula, we obtain 
\beq
A_{W}( B(k) \rightarrow D(p) \pi(k-p)) = i f_{\pi} \frac{G_{F}}{\sqrt{2}} 
V_{cb}V^{*}_{ud}(k^{2} - m_{D}^{2}) f_{+}(0, k^{2}) \; , 
\eeq
where the form factor $  f_{+}(0,k^{2}) $ is defined according to 
\beq
\ll D(p)| \bar{c} \Gamma_{\mu}b | B(k) \rl = f_{+}((k-p)^{2},k^{2})(k_{\mu} +
p_{\mu}) + f_{-}((k-p)^{2},k^{2})(k_{\mu} - p_{\mu})
\eeq
Here we have introduced an additional variable $ k^{2} $ into the standard 
definition of
the $ B \rightarrow D $ form factor bearing in mind a subsequent 
analytical continuation in 
$ k^{2} $. We note that on the mass shell 
the form factor $ f_{+}(0, m_{B}^{2}) $ can be expressed via the 
Isgur-Wise function \cite{IW}
\beq
f_{+}(0, m_{B}) = \frac{m_{B}+ m_{D}}{\sqrt{4 m_{B} m_{D}}} \xi(1 + 
\frac{(m_{B}-m_{D})^{2}}{2 m_{B} m_{D}})
\eeq
We will not use, however, this correspondence but rather derive a new 
light cone sum rule which is more suitable for an analytical 
continuation and refers 
directly to $ f_{+}(0,m_{B}^{2}) $ . We delay 
a corresponding calculation until the next section and concentrate 
first on the structure of the 
integral (1). An integration of Eq.(1) yields
\beq
Disc \; M(s) = \frac{i}{4 \pi} \cmb f_{+}(0,s) \int_{-1}^{1} 
d(\cos{\theta})
T(s,t ) \frac{(s-m_{D}^{2})^{2}}{4 s} \Theta(s - m_{D}^{2})
\eeq
Here $ \theta $ is an angle between the momenta $ \vec{k} - 
\vec{l} $ and 
$ \vec{p} $ :
\beq
t = (k-l-p)^{2} = -  \frac{(k^{2} - m_{D}^{2})^2}{2 k^{2}} (1 - 
\cos{\theta}) 
\eeq
The following observation is crucial for the rest of this paper. 
First, we
note that Eq.(6) contains the s-wave partial strong amplitude, in 
accord with the angular momentum
conservation. Therefore, one 
could think at the first sight that the Regge phenomenology, as 
being stemming from 
a resummation of high partial amplitudes, is of no use here. 
However, it has been known 
for a long time that partial amplitudes, obtained from 
Regge amplitudes by extrapolation into a low energy region, reveal resonance 
structures. This means that a correct answer is to be obtained by first
resummation of partial waves into a Regge form, and then re-expanding
it back into partial amplitudes.
It is useful for a moment to think of the final state
interaction as being proceeding 
through a resonances formation.
Anticipating the later result (see Eq.(13) below) we note that the 
discontinuity in the dispersion integral
is a slow (power-like) function of $ s $. Therefore in view of a 
large parameter $ m_{B}^{2} $ in the denominator a very 
large region of $ s $ from the threshold $ s = (m_{D} + m_{\pi})^{2} 
$ to $ s \sim m_{B}^{2} $ contributes in the 
dispersion integral. It is quite clear that in this parametrically 
large region many overlapping resonances are
created. According to the well known particle - reggeon duality 
phenomenon s-channel resonances can be be described 
by a t-channel reggeon exchange \cite{PDBC}. To avoid possible 
misunderstanding, we should emphasize a difference of our case from
the duality picture which applies, strictly speaking, only to an 
imaginary part of scattering amplitude averaged over some region 
in energies. Our consideration applies directly to the scattering
amplitude, as follows from the discussion above and Eq.(6). Thus, we 
conclude that the integral in Eq.(6) is dominated by a strong scattering 
amplitude in the Regge kinematics. Taking into account the leading $ 
\rho - $ meson trajectory, we obtain
\beq
T (s,t) =  \beta(t) \xi(t)\left( \frac{s}{s_{0}} \right)^{\alpha(t)}
\eeq
Here $ \alpha(t) = \alpha(0) + \alpha' t $ is the Regge trajectory, $ 
s_{0} $ is a scale parameter,  $  \xi(t) $ is the signature factor 
\beq
\xi(t) = \frac{ 1 - \exp{ [-i \pi \alpha(t)] }}{\sin{\pi \alpha(t)}}
\eeq
and $  \beta(t) $ stands for the reggeon residue. In what follows we 
adopt a model for the functional t-dependence
in $  \beta(t) $ obtained in dual models \cite{PDBC} and the quark-gluon 
string model \cite{Kaid} : 
\beq
\beta(t) = \frac{\beta(0)}{\Gamma [ \alpha(t)] }
\eeq
We next calculate the integral in Eq.(6)  
neglecting the t-dependence of the product $ 
\Gamma(\alpha(t)) \sin{\pi \alpha(t)} $ 
\beq
\Gamma(\alpha(t)) \sin{\pi \alpha(t)} \simeq \Gamma(\alpha(0)) \sin{\pi 
\alpha(0)} \simeq \sqrt{ \pi}
\eeq
and approximating  $ t \simeq - s/2 ( 1 - \cos{\theta} ) $ . Then
\bea
\int_{-1}^{1}d(\cos{\theta}) T(s,t) = \frac{2 \beta(0)}{ \sqrt{\pi} 
\alpha' s}  
\left( \frac{s}{s_{0}} \right)^{\alpha(0)}  \nonumber \\
\times \left[  \frac{1 - \exp{ [ - 
\alpha' s \ln{\frac{s}{s_{0}}} ]}}{ \ln{ \frac{s}{s_{0}} } }  
+ i \frac{ 1 - \exp{ [ - \alpha' s ( \ln{ \frac{s}{s_{0}} } - i \pi ) ] }}{
\ln{ \frac{s}{s_{0}} } - i \pi } \right]
\eea
Since the exponents are small in comparison to unity above the threshold 
$ s = m_{D}^{2} $, we finally obtain
\beq
Disc \; M(s)  \simeq  \frac{i}{8 \pi} \frac{G_{F}}{ \sqrt{2}} 
V_{cb}V_{ud}^{*} f_{\pi} f_{+}(0,s) F(s) \frac{ \beta(0)}{ \sqrt{ \pi} 
\alpha' s } \left( \frac{s}{s_{0}} \right)^{\alpha(0)} \Theta(s - m_{D}^{2})
\eeq
where
\beq
F(s) =  \frac{1}{ \ln{(2 \alpha' s)}}
+ \frac{i}{ \ln{(2 \alpha' s)} -
i \pi } 
\eeq 
In these formulae the intercept $ \alpha(0) \simeq 0.5 $ and the slope 
of the 
Regge trajectory $ \alpha' \simeq 1 \; Gev^{-2} $. Furthermore, it is 
important to stress that the scale parameter  $ s_{0} \equiv s_{0}^{\pi D}
$ is not constant, but varies with $ m_{c} $. To parametrize this 
dependence, we will use the relation obtained in the quark gluon string
model \cite{Kaid}
\beq
s_{0}^{\pi D} = s_{0}^{\pi \pi} \frac{ \bar{x}_{\pi}^{d}}{ \bar{x}_{D}^{d}}
\simeq ( \alpha')^{-1} \frac{m_{c}}{2 \Lambda}
\eeq
where $ \bar{x}_{\pi}^{d} \; ( \bar{x}_{D}^{d} ) $ is the average 
momentum fraction carried by the d-quark in the $\pi $- (D-) meson and 
$ \Lambda \simeq 300 - 400 \; MeV $ is the average virtuality of the light
quark in the D-meson. For the physical value $ m_{D} \simeq 1.9 \; GeV $ one
obtains $ s_{0}^{\pi D} \simeq 2 (\alpha')^{-1}$.
It is worth noticing that the standard dual model arguments would rather 
suggest a universal expression for the scale factor $ s_{0}^{\pi D}    
\simeq(\alpha')^{-1}$. 

\section{Light cone sum rule for the $ B \rightarrow D $ form factor}

Before proceeding to a calculation of the dispersion integral we need an
estimate for a functional $ s $ dependence of the $ B \rightarrow D $ 
transition form factor $ f_{+}(0, s)$ . The idea of a method presented in 
this section is borrowed from a very similar problem of the pion form factor 
at intermediate momentum transfers, which is most conveniently solved
by the light cone QCD sum rules method \cite{BH} (see also Ref. \cite{BKR}).
  We consider the following correlation function 

\beq
T_{\mu \nu}(p, k) = i \int dx e^{ikx} \ll D(p) | T \left\{ \bar{c} 
\gmmu b(x) \bar{b} \gmnu \gmf d(0) \right\} | 0 \rl
\eeq
for Minkowskian $ k^{2} = 0 $ and large Euclidean momentum  
$ (p+k)^2 \rightarrow - \infty $. 
The B-meson contribution to this correlation function is 
\beq
- 2 p_{\mu} k_{\nu} i f_{B} \frac{1}{m_{B}^{2} - (p+k)^2 } 
f_{+}(k^{2}) + \cdots \; \; , 
\eeq
where ellipses stand for other Lorentz structures. 
On the other hand, the
correlation function (12) can be calculated in QCD at large Euclidean
momenta $ ( p+ k)^{2} $. Indeed, in this case the virtuality of the 
heavy quark, which is of order $ m_{b}^{2}
- (p+k)^{2} $, is large. Thus, we can expand the heavy quark
propagator in powers of slowly varying fields residing in the D-meson,
which act as external fields on the propagating heavy quark. The
expansion in powers of an external field is also the expansion of the
propagator in powers of a deviation from the light cone $ x^{2} \simeq 0
$. The leading
contribution is obtained by using the free heavy quark propagator in the
correlation function (12). We obtain
\beq
T_{\mu \nu}(p,k) =  \int \frac{d^{4}x d^{4}l}{(2 \pi)^{4} ( m_{b}^{2}
- l^{2})} e^{i(k-l)x} \langle D(p) | \bar{c}(x)
\gmmu  (l_{\xi} \gamma_{\xi} + m_{b})  \gmnu \gmf d(0) | 0 \rangle
\eeq
 In this formula a
path-ordered gauge factor between the quark fields is implied, as
required by gauge invariance. In the particular case of the
Fock-Schwinger gauge $ x_{\mu} A_{\mu}(x) = 0 $ this
factor is equal to unity.
We see that the answer is expressed via the one-meson matrix element of
the gauge invariant non-local operator with a light-like separation $ x^{2}
\simeq 0 $. This matrix element defines a light cone D-meson distribution
amplitude. The leading twist distribution amplitude $ \phi (u) $ is 
introduced as follows \cite{CZ} :
\beq
\ll D(p) | \bar{c}(x) \gmmu \gmf d(0) | 0 \rl = - i f_{D} p_{\mu} 
\int_{0}^{1} du \; \exp{(iupx)} \phi(u) 
\eeq
Within this definition we obtain to the leading twist accuracy 
(hereafter $ 
\bar{u} = 1-u $) \beq
T_{\mu \nu}(p,k) = - i f_{D} p_{\mu} k_{\nu} \int_{0}^{1} du \;
\frac{ \phi(u)}{m_{b}^{2} + u \bar{u} m_{D}^{2} - u (p+k)^{2}}
 + \cdots
\eeq
To match Eq.(19) with the B-meson contribution (16) to the correlation
function (15), we note that Eq.(19) can be re-written as the dispersion 
integral with $ (m_{b}^{2} + u \bar{u} m_{D}^{2})/u $ being the mass of 
an intermediate state. The duality prescription tells that this invariant 
mass has to be restricted from above by the duality threshold $ s_{0}
= (m_{b} + m_{0})^{2} $, where $ m_{0} $ does not depend on $ m_{b} $. 
For physical value $ m_{b} \simeq 4.7 \; GeV $ one gets $ s_{0} \simeq 
35  \; GeV^{2} $, thus $ m_{0} \simeq 1.2 \; GeV $. We then obtain the 
following sum rule
\beq
\frac{2 f_{B}}{m_{B}^{2} - (p+q)^{2}} f_{+}(0,m_{B}) = 
f_{D} \int_{1-\frac{2m_{0}}{m_{b}}}^{1} du \; \frac{\phi(u)}{m_{b}^{2} + 
u \bar{u} m_{D}^{2} -u (p+q)^{2}} + \cdots
\eeq
To analyze a parametric behavior of the form factor $ f_{+}(0,m_{B}) $, 
we first note that as $ m_{b}^{2} \gg m_{c}^{2} $ we can neglect the 
second term in the denominator in the right hand side of Eq.(20). Next 
one has to distinguish between two different kinematical regimes : \\
(i) In the limit $ m_{b} \rightarrow \infty $ and $ m_{c} \rightarrow 0 $
D-meson becomes pion, the corresponding process being $ \bar{B} \rightarrow
\pi^{+} \pi^{-} \rightarrow 2 \pi^{0} $.  
In this case the asymptotic as $ u  \rightarrow 1 $ behavior of $ \phi(u) $
is $ \phi(u) \simeq 6 (1-u) $ and then
\beq
f_{+}(0,m_{B}) \simeq \frac{6 f_{\pi} m_{0}^{2}}{f_{B} m_{b}^{2}} \sim 
m_{B}^{-3/2} \eeq
(here we have used the fact that for $ m_{Q} \rightarrow \infty $ $ f_{Q}
\sim m_{Q}^{-1/2} $ up to logarithmical corrections.)
This dependence is well known from the light cone sum rules for the 
$ B \rightarrow \pi $ transition \cite{BKR, BBKR}. \\
(ii) More interesting is the case $ m_{c},m_{b} \rightarrow \infty $ with
$ m_{b} \gg m_{c} $. Then the D-meson distribution amplitude is 
strongly
peaked at equal velocities of quarks, i.e. concentrated near $ u =1 $ 
\cite{CZ}. We adopt the parametrization \cite{Br} 
\beq
\phi_{D}(u) = A_{D} \frac{ u^{2} \bar{u}^{2}}{ [ \varepsilon^{2} u +
\bar{u}^{2} ]^{2} } \; ,
\eeq
where $ \varepsilon = \Lambda /( \Lambda + m_{c} ) $ and $ A_{D} $ is 
fixed by the normalization ( $  A_{D} \simeq 0.67 $ for $ m_{c} \simeq 
1.3 \; GeV $ and $ \Lambda \simeq 0.33 \; GeV $ ). Using Eqs. (21) and 
(23), we obtain 
\beq
f_{+}(0,m_{B}) \simeq \frac{4 A_{D} f_{D} m_{0}^{3}}{ \left[ \varepsilon^{2}
+ \frac{4 m_{0}^{2}}{m_{B}^{2}} \right]^{2} } \frac{1}{ f_{B} m_{B}^{3}}
\equiv \xi  m_{B}^{-5/2}
\eeq
More faster in comparison with (22) fall off of the form factor $ 
f_{+}(0,m_{B}) $ indicated in Eq. (24) appears quite reasonable. 
As $ \phi_{D}(u) $ is strongly peaked near $ u = 1$, the integral 
in Eq. (21) becomes very sensitive to a change of the integration 
limit. Note in passing that this is precisely the reason why a naive
ansatz of the type 
 $ \phi(u) = \delta (u - 1 + \Lambda /m_{c} ) $ would produce 
completely wrong answer.

\section{Residue factor $ \beta(0) $ and $ g_{\rho D D}$  coupling constant}

In this section we obtain an estimate for the residue $ \beta(0) $. To this
end, we make use of the fact that a reggeon-particle vertex is an 
analytical continuation of the t-channnel resonance
coupling constant into the region $ t \sim 
0 $ \cite{PDBC}. 
We thus consider the charge exchange process $ D^{+} 
\pi^{-} \rightarrow D^{0} \pi^{0} $ proceeding for $ t \simeq 
m_{\rho}^{2} $ via the $ \rho $- meson exchange. The asymptotic ( for $ s 
\rightarrow \infty $ ) expression for the amplitude is 
\beq
T(s,t) = g_{\rho \pi \pi} g_{\rho D D} \frac{2 s }{t -m_{\rho}^{2}}
\eeq
The $ \rho - \pi $ vertex is well known:
\beq
g_{\rho \pi \pi} \simeq g_{\rho } \; \; , \; \; 
\ll 0 | e_{u} \bar{u} \gmmu u + e_{d} \bar{d} \gmmu d  | \rho \rl = 
\frac{m_{\rho}^{2}}{g_{\rho}} \varepsilon_{\mu} \; , 
\eeq
( $ \varepsilon_{\mu} $ stands for the $ \rho $- meson polarization 
vector) and 
thus we have to estimate only the $ \rho D $ vertex $ g_{\rho D D} $.
Here we again use the light cone QCD sum rules technique \cite{BF}. Consider 
the correlation function 
\beq
T_{\mu \nu }(p,q) = i \int dx \; e^{ipx} \ll \rho(q) | T \{ \bar{d} \gmmu 
\gmf c(x) \bar{c} \gmnu \gmf u(0) \} | 0 \rl
\eeq
The contribution of interest is 
\beq
p_{\mu}( p_{\nu} + q_{\nu}) ( \varepsilon p) g_{\rho D D} f_{D}^{2} \;  
\frac{1}{ ( m_{D}^{2} - p^{2}) ( m_{D}^{2} - (p+q)^{2})}
\eeq
On the other hand, we calculate this correlation function at large 
Euclidean momenta $ p^{2} , (p+q)^{2} \rightarrow - \infty $ using the 
Operator   
Product Expansion near the light cone \cite{BF, BBKR}. To the leading twist 
accuracy we obtain
\beq
T_{\mu \nu} = \frac{1}{2} p_{\mu} q_{\nu} ( \varepsilon p) 
 f_{\rho} m_{\rho} \int_{0}^{1} \; du \frac{g_{\perp}^{(a)} (u) - 4  
\Phi_{\parallel}(u)}{ [ m_{c}^{2} - (p+uq)^{2} ]^{2} } \; , 
\eeq
where the twist 2 distribution amplitudes $ g_{\perp}^{(a)}(u) $ and $ 
\Phi_{\parallel}(u) $ are defined as follows \cite{CZ, BBl}:
\bea
\ll \rho(q) | \bar{d}(x) \gmmu u(0) | 0 \rl &=&  q_{\mu} ( \varepsilon x)
 f_{\rho} m_{\rho} \int_{0}^{1} \; du e^{iqx} \Phi_{\parallel}(u) 
\nonumber \\
 &+& \varepsilon_{\mu} f_{\rho} m_{\rho}\int_{0}^{1} \; du e^{iqx}
g_{\perp}^{(v)}(u)
\eea

\beq
\ll \rho(q) | \bar{d}(x) \gmmu \gmf  u(0) | 0 \rl = - \frac{1}{4}f_{\rho} 
m_{\rho} \varepsilon_{ \mu \nu \lambda \sigma} \varepsilon_{\nu} q_{\lambda}
x_{\sigma} \int_{0}^{1} \; du e^{iqx} g_{\perp}^{(a)}(u)
\eeq
To the lowest conformal spin accuracy the distribution amplitudes have very
simple forms \cite{BBl} : 
\bea
\Phi_{\parallel}(u) &=& i \frac{3}{2} u \bar{u} (2u -1) \nonumber \\
g_{\perp}^{(a)}(u)  &=& 6 u \bar{u}
\eea
We next use the identity $ (p+uq)^{2} = \bar{u} p_{1}^{2} + u p_{2}^{2} - u 
\bar{u}
q^{2} $ with $ p_{1}^{2} = p^{2} $ and $ p_{2}^{2} = (p+q)^{2} $ and the 
formula for the double Borel transformation in respect to  $ p_{1}^{2}  $ 
and $ p_{2}^{2} $
\beq
B_{M_{1}^{2},M_{2}^{2}} \frac{\Gamma(\nu)}{ [ m_{c}^{2} + u \bar{u}q^{2}
- \bar{u}p_{1}^{2} - u p_{2}^{2} ] ^{\nu} } = (M^{2})^{2-\nu} \exp{ [ - 
\frac{m_{c}^{2}}{M^{2}} - \frac{q^{2}}{M_{1}^{2} + M_{2}^{2}} ] } \delta(
u - u_{0})
\eeq
where 
\beq
M^{2} = \frac{M_{1}^{2}M_{2}^{2}}{M_{1}^{2} + M_{2}^{2}} \; \; , \; \; 
u_{0} = \frac{M_{1}}{M_{1}^{2} + M_{2}^{2}}
\eeq
In a final expression we set $ M_{1}^{2} = M_{2}^{2} = 2 M^{2} $ and account
for the fact that $ \Phi_{\parallel}(1/2) = 0 $. We then arrive at a 
simple sum rule
\beq
g_{\rho DD} \simeq \frac{f_{\rho}m_{\rho}}{2 f_{D}^{2}} 
g_{\perp}^{(a)}(\frac{1}{2}) \left( \exp{(-\frac{m_{c}^{2}}{M^{2}} - 
\frac{m_{\rho}^{2}}{4 m^{2}}) } - \exp{(- \frac{s_{0}}{M^{2}}) } \right)
\eeq
Here the second exponent corresponds to a continuum subtraction with 
$ s_{0} \simeq 6 \; GeV^{2} $. This sum rule yields
\beq
g_{\rho DD} \simeq 4
\eeq
On the other hand, the Regge formula taken for $ t > 0 $ provides
\beq
T(s, t \simeq m_{\rho}^{2}) \simeq \frac{\beta(m_{\rho}^{2})}{\pi} \frac{s}{
t -  m_{\rho}^{2}}
\eeq 
Comparing these expressions and taking into accout Eq.(10), we arrive at
an estimate
\beq
\beta(0) \simeq 2 \pi \Gamma[ \alpha(m_{\rho}^{2}) ] g_{ \rho \pi \pi}
 g_{\rho DD} \simeq 140
\eeq

\section{Reggeon versus direct amplitude}

To obtain the amplitude of the process of interest, we write down  
an unsubtracted dispersion relation 
\beq
M (s) = \frac{1}{2 \pi i} \int_{m_{D}^{2}}^{\infty} \; d s' \; 
\frac{ Disc \; M(s')}{ s' - s - i \varepsilon}
\eeq
 To calculate the integral, we use a 
standard trick. The integral is first calculated for negative $ s = - 
m_{B}^{2} $  and then one makes the analytical continuation 
$ m_{B}^{2} \rightarrow  m_{B}^{2} \exp{(-i\pi )} $ to positive $
s = m_{B}^{2} $. To make subsequent formulas feasible, we neglect in 
this calculation 
the s-dependence of the function $ F(m_{B}) $ (see Eq. (14)). It is 
convenient to parametrize this quantity in the form
\beq
F(m_{B}) = |F| \exp{(i \phi)}
\eeq
with ( $ a = \ln{(2 \alpha' m_{B}^{2})} $ )
\bea
|F| &=& \frac{1}{a} \left( \frac{2 a^{2} + \pi^{2} - 2 \pi 
a}{ \pi^{2} + a^{2}} \right)^{ \frac{1}{2} }  \simeq 0.2 \nonumber \\
\phi &=&  atan \frac{a^{2}}{a^{2} + \pi^{2} - \pi a } \simeq 50^{o} 
\eea
(here numerical values refer to the physical mass $ m_{b} \simeq 4.7 
\; GeV $). Again, we distinguish between two different limiting cases:
\\
(i) $ m_{b} \rightarrow \infty $ and $ m_{c} \rightarrow 0 $. The 
corresponding physical process is $ \bar{B} \rightarrow \pi^{+} 
\pi^{-} \rightarrow 2 \pi^{0} $. In this case the scale parameter 
is the standard $ s_{0}^{ \pi \pi} = (\alpha')^{-1} $. The 
asymptotic with $ m_{b} \rightarrow \infty $ behavior of the $ B 
\rightarrow \pi $ transition form factor is given (in the chiral 
limit) by Eq. (22). However, the asympotic estimate (22) cannot be 
analytically continued down to the threshold value $ s = 0$. To 
get an interpolating formula, we note that at $ m_{b} \rightarrow 
0 $ $ f_{+}(0,m_{B}) $ reduces to the pion electomagnetic form factor
which equals unity at zero momentum transfer. We model this by 
the interpolating formula which is approximately valid for all $ m_{B} $
in the dispersion integral
\beq
f_{+}(0,s) = \frac{6 f_{\pi} m_{0}^{2}}{ f_{B} s + 6 f_{\pi} 
m_{0}^{2}}
\eeq
Neglecting for simplicity the $ m_{B} $-dependence of $ f_{B} $ 
during the calculation of the integral, we finally obtain
\beq
M_{\pi \pi}^{(r)} (m_{B}) = \frac{G_{F}}{ \sqrt{2}} V_{ub}V_{ud}^{*}
\frac{3\beta(0) |F|}{8 \pi\sqrt{\pi (\alpha')^{3}}} \frac{m_{0}^{2}
f_{\pi}^{2}}{ m_{B} f_{B}} e^{i(\phi + \frac{\pi}{4} )} \sim 
m_{B}^{-1/2}
\eeq
(here the subscript "r" means the reggeon-exchange induced 
mechanism of
weak decay.) \\
(ii) $ m_{c},m_{b} \rightarrow \infty $ with $ m_{b} \gg m_{c} $. 
The corresponding physical process is $ \bar{B} \rightarrow D^{+}
\pi^{-} \rightarrow D^{0} \pi^{0} $. Using Eqs. (15), (24) and 
(38), we obtain in this case
\beq
M_{D\pi}^{(r)} (m_{B}) =  \frac{G_{F}}{ \sqrt{2}} V_{cb}V_{ud}^{*} 
f_{\pi}
\frac{ \beta(0)}{ 8 \pi \sqrt{ \pi \alpha'm_{c}} } \xi |F| 
m_{B}^{-3/2} e^{i 
(\phi + 3 \pi/4 - 2 \eta/ \pi )}  \left(1  - \frac{2 \eta}{ \pi} 
 \right) 
\eeq
In deriving this formula we have restricted ourselves by the 
linear order in $ \eta \equiv \sqrt{m_{D}/m_{B}} $. 
The answers  we have found are to be compared with perturbative QCD  
amplitudes  of the direct decays which are induced by the well known 
effective non-leptonic Hamiltonian (we display only the Cabibbo favored part)
\beq
H_{eff} = \frac{G_{F}}{\sqrt{2}} 
V_{cb}V_{ud}^{*} [ C_{1}(\mu) O_{1}^{u} + C_{2} 
(\mu)
O_{2}^{u} ] 
\eeq
where
\beq
O_{1}^{u} = (\bar{c} \Gamma_{\mu} b ) ( \bar{d} \Gamma_{\mu} u ) \; \; and
\; \; O_{2}^{u} = ( \bar{d}\Gamma_{\mu} b )( \bar{c} \Gamma_{\mu} u)
\eeq
The Wilson
coefficients $ C_{i}( \mu) $ are due to the renormalization of the bare
Hamiltonian $ H_{W} \sim O_{1} $ by hard gluons with virtualities larger
than $ \mu^{2} = O( m_{b}^{2} ) $. In the leading-log approximation
with $ \mu \simeq 5 \; GeV $
 ,  $
n_{fl} = 5 $ and  $ \Lambda_{\bar{MS}} \simeq 200 \; MeV $ \cite{AM}
\beq
C_{1} = 1.117 \; \; and \; \; C_{2} = - 0.266
\eeq
For the $ \bar{B}^{0} \rightarrow D^{0} \pi^{0} $ decay $ O_{2} $ 
factorizes and
\bea
\langle D^{0} \pi^{0} | H_{eff} | \bar{B}^{0} \rangle \sim ( C_{2} + \frac{
C_{1}}{N_{c}} ) \langle D| \bar{c} \Gamma_{\mu} u | 0 \rangle \langle \pi
| \bar{d} \Gamma_{\mu}b | B \rangle \nonumber \\
 + C_{1} \langle D \pi | 2 ( \bar{c} \Gamma_{\mu} t^{a} u ) ( \bar{d}
\Gamma_{\mu} t^{a} b ) | B \rangle
\eea
The non-factorizable contribution to this decay
which is given by the second matrix element in this formula has been 
estimated in Ref. \cite{me} by the light cone QCD sum rule method. There has 
been found  the following estimate for the ratio of
 the non-factorizable to the factorizable $ 1/N_{c} $
amplitudes :
\beq
r \equiv \frac{M^{(nf)}}{M^{(f)}} \simeq - \frac{N_{c}}{4 \pi^{2} 
f_{D}^{2}}
\frac{ p_{\alpha} \ll \pi | \bar{d} \tilde{G}_{\alpha \mu} \gmmu \gmf b |
B \rl }{ ( m_{B}^{2} - m_{D}^{2}) f_{\pi}^{+}(m_{D}^{2})} \simeq - 0.7
\eeq
where the the $ B \rightarrow \pi $ transition form factor $ f_{\pi}^{+}
$ is defined as in Eq.(4).
The direct decay amplitude is then
\beq
M^{(dir)}(m_{B}^{2}) = i \frac{G_{F}}{\sqrt{2}}V_{cb}V_{cs}^{*}
\left[ C_{2}
 +\frac{C_{1}}{N_{c}}(1 + r) \right] f_{D} f_{\pi}^{+}(m_{D}^{2}) 
(m_{B}^{2} - m_{D}^{2} ) \sim m_{B}^{1/2}
\eeq
We therefore see that the reggeon-induced amplitude is suppressed as $ 
1/m_{B}^{2} $ in respect to the direct one, while at $ m_{B} = 5.27 \; 
GeV $ one obtains 
\beq
\frac{M_{D\pi}^{(r)}}{M_{D \pi}^{(dir)}} \simeq - \frac{0.09}{C_{2} 
+\frac{C_{1}}{N_{c}}(1 + r)} e^{i (\phi + \pi/4 - 2 \eta/ \pi )} \simeq 0.18
+ 0.58 i  
\eeq
The last equality in this formula is obtained with the particular value (49).
Repeating the same procedure yields a similar formula for  
the $ B \rightarrow 2 \pi^{0} $ 
decay. In this case (if one neglects a penguin contribution) the amplitude 
factorizes analogously to (48), but this time the suppression is only $ 
1/m_{B} $. Numerically one finds in this case
\beq
\frac{M_{\pi\pi}^{(r)}}{M_{\pi\pi}^{(dir)}} \simeq \frac{0.12}{C_{2} + \frac{
C_{1}}{N_{c}}(1+r)} e^{i(\phi - \pi/4)} \; , 
\eeq
where the parameter $ r $ is defined analogously to Eq.(49).  
Our final results Eqs.(51) and (52) suggest that the
rescattering amplitudes  are not small and must be added to the 
amplitudes 
of direct decays. In this sense the latter can be thought of as radiative
corrections to "tree" level amplitudes induced by soft FSI. We conclude that
perturbative QCD most likely does not exhaust non-leptonic B-decays. The 
physical b-quark mass, albeit large, is still beyond the onset of the 
asymptotic QCD regime in nonleptonic heavy meson decays. An example of a 
similar interplay between soft and hard QCD is discussed in the next 
section.

   We would like to end up this section with a comment on a relation 
between the direct and rescattering amplitudes. For definiteness, we will 
focus on  the QCD sum rule approach. As 
has been explained in the introduction, any information about the 
imaginary part of the amplitude is lost there. Our method therefore yields 
the entire phase of the decay. The issue of the real part of the 
rescattering amplitude is slightly more subtle. Real parts of the amplitudes 
calculated in this paper correspond to the so-called "parasite" 
contributions to the sum rules (cf. Ref. \cite{BS87}) and are due to a 
possibility to produce states below the B-meson pole in the corresponding
channel. 
If the latter are 
omitted from an analysis of QCD sum rules (this require a careful 
analysis of corresponding Operator Product Expansion), then the two 
amplitides are simply added.  

\section{Concluding remarks}

In this paper we presented a semi-quantitative calculation of FSI effects 
in color suppressed B-decays. Our results suggest a rather dissapointing 
conclusion that the soft physics prevents currently used methods from 
reaching an accuracy better than 30 -50 \% , the error is mainly due to a 
poor knowledge of multiparticle scattering FSI. Effects of soft FSI are 
as important as perturbative (and non-pertrurbative) QCD corrections up to 
the physical value of the b-quark mass, though they are parametrically 
suppressed in agreement with the color transparency (CT) picture of 
Bjorken \cite{BJ}. In a sense the situation is reminiscent of the case
of the pion electromagnetic form factor, which is dominated by the soft 
physics up to currently available momentum transfers $ Q^{2} \simeq 10 \; 
GeV^{2} $ \cite{BH}. There an additional suppression $ 1/Q^{2} $ of the 
soft contribution overcomes the leading hard contribution due to a large 
numerical factor. It may yet seem surprising that in our problem 
perturbative QCD is insufficient even at much higher momenta $ \sim 
m_{B}^{2} $. 

   A similar to our approach to the analysis of FSI effects in B-decay 
has been suggested in a recent paper \cite{Don} (see also \cite{Zh}). 
We differ from Refs. \cite{Don, Zh} in two aspects. We concentrate on 
FSI reducible to the leading reggeon exchange, while the authors of 
\cite{Don, Zh} study worse understood elastic rescattering effects. 
Secondly, we calculate not only the phase but also corresponding 
corrections to the real parts of amplitudes. What if we apply the 
formulae of proceeding sections to the pomeron amplitude of elastic 
rescattering? We have checked that in this case we reproduce the 
conclusions of \cite{Don}, i.e. the corresponding amplitude is not 
suppressed respectively to the direct one. Along with the two channel 
model used in \cite{Don} multiparticle contributions do not seem to 
cancel the parametric $ m_{B}^{1/2} $ dependence of the pomeron 
amplitude. It appears therefore that the pomeron breaks down the CT 
picture of Ref \cite{BJ}. We would like to mention, however, that 
assumptions underlying this conclusion are different from those done in 
the CT scenario, thus the final conclusion of Ref. \cite{Don} does not 
come too surprising. We feel that a space time picture of B-decays 
similar to those applied to high energy scattering would be desirable for 
a further study of this issue.

\section{Acknowledgments}   
 
 We would like to thank G. Eilam, A. Gotsman, M. Gronau, U. Maor and S. 
Nussinov for useful conversations and interest in this work. B.B. is 
grateful to M. Shifman for discussions. I.H. would like to especially 
acknowledge stimulating discussions with L. Frankfurt on a number of 
subjects related to this work.  
We are grateful to Da-Xin Zhang for bringing Ref. 
\cite{Zh} to our attention.

\end{document}